\documentclass[conference,a4paper]{IEEEtran}

\usepackage{amsmath}
\usepackage{amssymb}
\usepackage{wasysym}
\usepackage{graphicx}
\usepackage{subfigure}
\usepackage{color}
\usepackage{bm,cite}
\usepackage{tikz}
\usepackage{mathrsfs}

\title{Non-Orthogonal Multiple Access for Degraded Broadcast Channels:
RA-CEMA}

\author{\IEEEauthorblockN{Alberto~G.~Perotti and Branislav~M.~Popovi\'c}
\IEEEauthorblockA{Huawei Technologies Sweden AB \\
Skalholtsgatan 9-11, SE--164 94 Kista, Sweden\\
E-mail: \texttt{\small\{alberto.perotti, branislav.popovic\}@huawei.com}}}

\begin{document}

\maketitle

\begin{abstract}
A new non-orthogonal multiple access scheme performing
simultaneous transmission to multiple users characterized by different
signal-to-noise ratios is proposed.
Different users are multiplexed by storing their codewords into
a multiplexing matrix according to properly designed patterns
and then mapping the columns of the matrix onto the symbols
of a higher-order constellation.
At the receiver, an interference cancellation algorithm
is employed in order to achieve a higher spectral efficiency than
orthogonal user multiplexing.
Rate-Adaptive Constellation Expansion Multiple Access (RA-CEMA)
is an alternative to conventional superposition coding as a solution
for transmission on the degraded broadcast channel. 
It combines the benefits of an increased
spectral efficiency with the advantages of reusing the coding and
modulation schemes already used in contemporary communication
systems, thereby facilitating its adoption in standards.
\end{abstract}

\section{Introduction}

Future wireless networks are expected to support significantly
increased Down-Link (DL) data traffic, either in the form of an increased number of User Equipments (UEs)
connected to each single DL transmitter (for example, the UEs might be sensors),
or in the form of multiple,
virtually concurrent data streams transmitted to the same UE
(for example, where each stream is delivered to a different
application running on the same UE).
Both cases can be modeled as an increased
number of high-rate DL data streams, which might be difficult or impossible to
support using orthogonal Multiple Access (MA) schemes.
The simultaneous transmission of multiple signals using some
common Resource Elements (REs) is the basic feature of
OverLoaded MA (OLMA) methods~\cite{bib:Ping06VTMag, bib:Saito13VTC,bib:Choi14ComLet}.
Practical OLMA schemes can be designed
starting from different scenarios, each one characterized by a
specific optimization criterion or target for the selection of transmission
parameters, leading to quite different solutions.
However, all OLMA
schemes have to ensure reliable separation/detection and
decoding of each multiplexed stream at the  intended UEs.

In one scenario, the optimization target is the maximization of the
aggregate DL spectral efficiency by simultaneous transmission to UEs
experiencing similar physical communication channel qualities.
The UEs that report to the transmitter similar Channel Quality
Indicators (CQI)\footnote{In LTE, each UE reports Channel Quality Indicators
to its serving base station. Typically, these indicators are related to the SINR
experienced by the UE and are used by the scheduler to select
transmission parameters.}
are grouped into the same category, and then
served by the same transmission resources when the instantaneous
channel conditions are the best at these resources.
The corresponding OLMA methods thus
preserve the same data
rate, the same transmitted energy per bit of each multiplexed stream,
and the
same scheduler design as if each of the streams would have been
transmitted alone on the observed time-frequency-space resources.
It further means
that the transmitted power per RE is increased proportionally to the overloading factor, i.e. the number of multiplexed streams.
The OLMA schemes designed using
this principle include, for example, Low-Density
Spread Multiple Access (LDSMA)~\cite{bib:Choi04ISSSTA, bib:BeekP09Globecom, bib:Popov14WCNC},
Enhanced/Turbo Trellis-Coded Multiple Access (ETCMA,
TTCMA)~\cite{bib:Perot14ITW, bib:Perot14GCws1}
and Constellation Expansion Multiple Access (CEMA)~\cite{bib:Perot14GCws2}.

In another overloading scenario, the target is to increase the number
of UEs served per RE, but without increasing the average transmitted
power.
The direct
consequence of conserving the transmitted power is that the achievable data
rates of each of multiplexed UE signals are lower than if each of them would
have been transmitted separately.
An additional target is to perform multiplexing in such a way that the aggregate rate
of the concurrently served UEs is larger than the aggregate rate  that can be
obtained by time sharing (time division) multiplexing of these UEs (where
each transmission
interval is split into sub-intervals corresponding to different UEs). It can be
shown that this target can be achieved only if the received Signal-to-Noise
Ratios (SNRs) of the multiplexed UEs are not equal.
It should be noted that this target is not equivalent to maximizing the
aggregate data rate per RE, as it can be shown that the aggregate data rate
cannot be larger than the maximum single UE data rate (obtained for
the UE with the highest received SNR).
Such a DL transmission to users with significantly different SNRs
is known
in information theory as
\emph{degraded} Broadcast Channel
(BC)~\cite{bib:CoverThomas, bib:ElGamalKim}.

A practical OLMA scheme for the degraded
BC is based on the amplitude-weighted superposition of 
modulated codewords for (typically two) different UEs.
UE-specific scaling coefficients are chosen with the constraint
of maintaining the total transmitted power equal to the average
power for single UE transmission~\cite{bib:Vanka12TranWC}.
We shall refer to such scheme as 
Amplitude-Weighted Non Orthogonal Multiple Access (AW-NOMA).
Scaling coefficients are changed during transmission in order to make the
system adaptive to the varying SNRs and data rate requirements of the
served UEs.

In the case of AW-NOMA, the transmitted signal is a series of
new constellation symbols, obtained by weighted sum of 
two conventional modulation symbols. The minimum 
Euclidean Distance (ED) of the new constellation symbols might 
be much smaller than the ED of the 
corresponding conventional constellation having the same 
asymptotic spectral efficiency. Smaller minimum ED 
of AW-NOMA constellation symbols might require 
smaller maximum allowed distortion in the transmitter 
hardware than currently specified by LTE standard through 
the requirements on maximum Error Vector Magnitude 
(EVM) for each modulation format. As the EVM 
requirements are specifications of the minimum 
implementation quality of the equipment to fully achieve 
potential gains of each supported constellation, it follows 
that if NOMA constellation for some power ratio of 
multiplexed UEs is different from already existing LTE 
modulation formats, it would directly demand a new 
standardization effort on the specification of the 
corresponding EVM requirements. This standardization 
effort is a separate problem from the actual EVM 
requirements which would result from it, as it consumes 
significant time and resources regardless of the possibility 
that in some cases the NOMA EVM requirements turn out 
to be the same as some already existing EVM requirements. 
Save a side a realistic possibility that the existing LTE EVM 
specifications, specifying the maximum allowed signal 
distortion introduced by transmitter hardware to the 
transmitted signal, might be too loose for superposed 
NOMA constellation symbols. 

The above potential standardization problems of AW-
NOMA were the major motive to develop an alternative 
scheme, which we called Rate-Adaptive Constellation 
Expansion Multiple Access (RA-CEMA). This new scheme, 
which will be discussed in the sequel, performs 
multiplexing of several coded data streams over a 
multiplexing matrix matched to the size of codewords, 
whose columns are then mapped to symbols of an expanded 
conventional constellation. This scheme can be considered 
as a generalisation of Bit Division Multiplexing (BDM)
scheme ~\cite{bib:SongTVT14}\footnote{BDM is actually
not an OLMA scheme, because the modulation format is fixed in 
advance depending on system targets such as coverage 
area etc., i.e. there is no scheduling. Besides, 
multiplexing is done similarly as in hierarchical
modulation~\cite{bib:dvb-t}, within a certain number of
symbols, much smaller than the codeword lengths.}.

The paper is organized as follows: 
Sec.~\ref{sec:sys-model} presents the model of transmission system herein
considered, Sec.~\ref{sec:ra-cema} describes the proposed OLMA scheme,
Sec.~\ref{sec:perf-eval} presents performance evaluation results
and Sec.~\ref{sec:conclusions} draws the final conclusions.

\begin{figure}[!ht]
\centering
\resizebox{.75\hsize}{!}{
\begin{tikzpicture}[scale=0.95]

\draw[->,line width=1pt]  (-.5,.75) -- (0,.75);
\draw[->,line width=1pt]  (-.5,.25) -- (0,.25);
\node[anchor=south] at (-.5,.75) {$\mathbf{b}_{\rm N}$};
\node[anchor=north] at (-.5,.25) {$\mathbf{b}_{\rm F}$};
\draw[line width=1pt]  (0,0) rectangle (2,1);
\node[text centered] at (1,.5) {\small\textsf{Transmitter}};
\node at (2.25,.25) {$\mathbf{x}$};

\draw[line width=1pt] (2,.5) -- (2.5,.5) -- (2.5,1.) -- (2.3,1.3) -- (2.5,1.) -- (2.7, 1.3);

\draw[line width=1pt]  (5.75,1) rectangle (8.25,2);
\node[text width = 1.5cm, text centered] at (7,1.5) {\bf\textsf{Far receiver}};
\node[anchor=north] at (5.5,1.5) {$\mathbf{y}_{\rm F}$};

\draw[->,line width=1pt]  (8.25,1.5) -- (8.75,1.5);
\node[anchor=south] at (8.75,1.5) {$\mathbf{\hat b}_{\rm F}$};
\draw[line width=1pt] (5.75,1.5) -- (5.25,1.5) -- (5.25,2.) -- (5.05,2.3) -- (5.25,2.) -- (5.45, 2.3);

\draw[line width=1pt]  (4,-3.5) rectangle (9.75,0.375);
\node[text centered] at (6.875, 0.125) {\bf\textsf{Near receiver}};

\draw[line width=1pt]  (6.75,-3.25) rectangle (9.25,-2.);
\node[text width = 2.cm, text centered] at (8,-2.625)
{\small\textsf{Near code-word decoder}};
\node[anchor=north] at (3.625,-.75) {$\mathbf{y}_{\rm N}$};

\draw[line width=1pt]  (6.75,-1.25) rectangle (9.25,-.25);
\node[text width = 2.cm, text centered] at (8.,-.75)
{\small\textsf{Far codeword decoder}};

\draw[line width=1pt]  (4.125,-2.375) rectangle (6.375,-1.);
\node[text width = 2.cm, text centered] at (5.25,-1.75)
{\small\textsf{Far codeword interference cancellation}};

\draw[->, line width=1pt]  (5.25,-.75) -- (5.25,-1.);
\draw[->, line width=1pt]  (8.,-1.25) -- (8,-1.75) -- (6.375, -1.75) ;
\draw[->, line width=1pt]  (5.25,-2.25) -- (5.25,-2.75) -- (6.75, -2.75) ;

\draw[->,line width=1pt]  (9.25,-2.5) -- (10.25,-2.5);
\node[anchor=south] at (10.25,-2.5) {$\mathbf{\hat b}_{\rm N}$};

\draw[<-,line width=1pt] (6.75,-.75) -- (3.5,-.75) -- (3.5,0.) -- (3.3,.3) -- (3.5,0.) -- (3.7, .3);
\node[anchor=north] at (3.625,-.75) {$\mathbf{y}_{\rm N}$};

\draw[->,line width=1pt, dashed]  (2.875,1.375) -- (5.,2.125);
\draw[->,line width=1pt, dashed]  (2.875,1.125) -- (3.25,.5);
\node at (3.625,2.) {$h_{\rm F}$};
\node at (3.375,.875) {$h_{\rm N}$};

\end{tikzpicture}
}
\caption{System model.}
\label{fig:sys-model}
\end{figure}
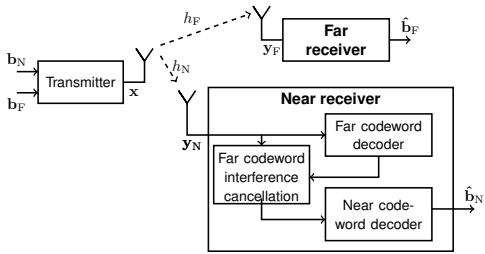

\section{System Model}
\label{sec:sys-model}

The transmission system considered in this paper is shown in Fig.~\ref{fig:sys-model}.
It consists of a transmitter, a \emph{far user} receiver and a \emph{near user} receiver.
The transmitter wishes to serve both users simultaneously by transmitting
information words \mbox{$\mathbf{b}_{\rm N} =(b_{{\rm N},0}, \ldots, b_{{\rm N},K_{\rm N}-1})$} and
$\mathbf{b}_{\rm F} = (b_{{\rm F},0}, \ldots, b_{{\rm F},K_{\rm F}-1})$ with the maximum possible data rates.

The channel from transmitter to near (resp. far) receiver is modeled as a
complex coefficient $h_{\rm N}$ (resp. $h_{\rm F}$) representing the combined
effect of propagation path loss, shadow fading and fast fading.
We assume that both receivers have perfect channel knowledge and that
$|h_{\rm N}| > |h_{\rm F}|$.

The received signal is
\begin{equation*}
\mathbf{y}_{u} = h_u \mathbf{x} + \mathbf{w}_{u}
\end{equation*}
where $u\in\{\rm N, F\}$ is the user index and $\mathbf{w}_{u}$ is a vector
representing additive white Gaussian noise (AWGN) whose elements are circularly
symmetric, zero-mean \emph{iid} Gaussian random variables with variance
$\sigma_{\rm w} = N_0/2$. Here, $N_0$ is the two-sided power spectral
density of noise.

The elements of $\mathbf{x}$ are uniformly drawn from a unit-energy constellation
and  the SNR of user $u$ is $\rho_u =|h_u|^2/N_0$.

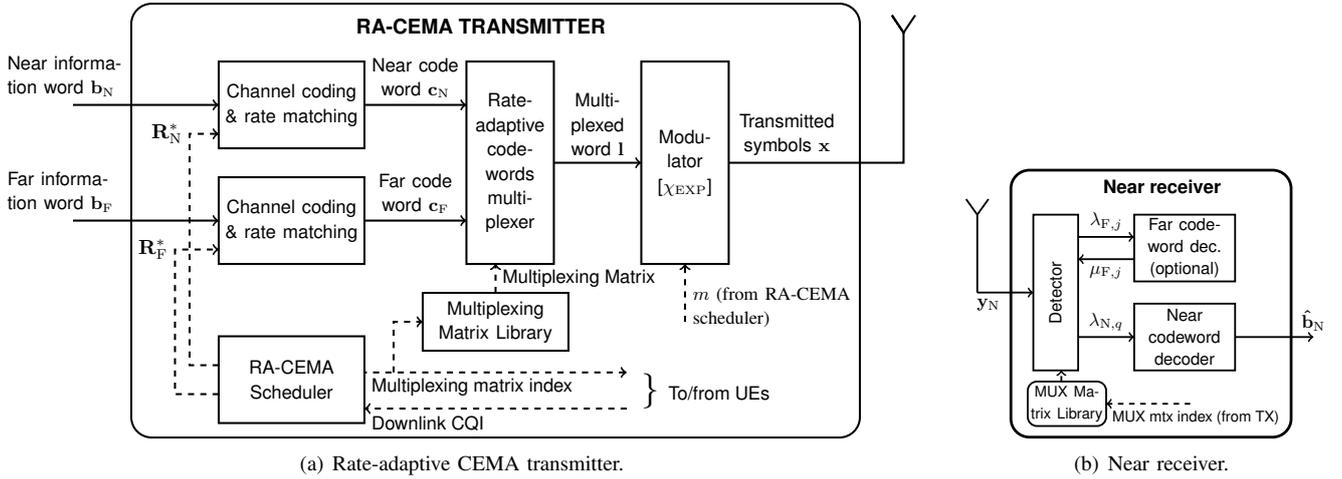
\begin{figure*}[!ht]
\centering
\subfigure[][Rate-adaptive CEMA transmitter.]{
\resizebox{.68\hsize}{!}{
\begin{tikzpicture}

\draw[line width=1pt, rounded corners=10pt]  (.5,0) rectangle (13,7.5);
\node[anchor=north, text centered] at (6.5,7.375) {\bf\textsf{RA-CEMA TRANSMITTER}};

\node[text width=2.2cm,anchor=west] at (-1.75,6.25) {\small\sf Near information word $\mathbf{b}_{\rm N}$};
\draw[->,line width=1pt]  (-.5,5.75) -- (2.,5.75);
\draw[line width=1pt]  (2.,5) rectangle (4.5,6.5);
\node[text width=2.3cm,text centered] at (3.25,5.75) {\small\sf Channel coding \& rate matching};
\draw[->,line width=1pt]  (4.5,5.75) -- (6.25,5.75);
\node[text width=1.5cm,anchor=south, text centered] at (5.375,5.75) {\small\sf Near code word $\mathbf{c}_{\rm N}$};

\draw[->,line width=1pt]  (-.5,3.75) -- (2.,3.75);
\draw[line width=1pt]  (2.,3) rectangle (4.5,4.5);
\node[text width=2.3cm,text centered] at (3.25,3.75) {\small\sf Channel coding \& rate matching};
\draw[->,line width=1pt]  (4.5,3.75) -- (6.25,3.75);
\node[text width=2cm,anchor=west] at (-1.75,4.25) {\small\sf Far information word $\mathbf{b}_{\rm F}$};
\node[text width=1.5cm,anchor=south, text centered] at (5.375,3.75) {\small\sf Far code word $\mathbf{c}_{\rm F}$};

\draw[line width=1pt]  (6.25,3.) rectangle (7.75,6.5);
\node[text width=1.4cm,text centered] at (7.0,4.75) {\small\sf Rate-adaptive code-words multiplexer};
\draw[->,line width=1pt]  (7.75,4.75) -- (9.25,4.75);
\node[text width=1.5cm,anchor=south, text centered] at (8.5,4.75) {\small\sf Multi-plexed word $\mathbf{l}$};

\draw[line width=1pt]  (9.25,3) rectangle (10.75,6.5);
\node[text width=1.4cm,text centered] at (10.,4.75) {\small\sf Modu-lator [$\chi_{\rm EXP}$]};
\node[text width=2.5cm,anchor=south, text centered] at (11.75,4.75) {\small\sf Transmitted symbols $\mathbf{x}$};
\draw[->,line width=1pt, dashed] (10,2.) -- (10,3.) node [near start,
right, text width=2.9cm] {\small $m$ (from RA-CEMA scheduler)};

\draw[line width=1pt]  (2.,0.25) rectangle (4.5,1.75);
\node[text width=1.8cm,text centered] at (3.25,1.) {\small\sf RA-CEMA Scheduler};
\draw[->,line width=1pt, dashed] (9,.5) -- (4.5,.5);
\node[anchor=west] at (4.5,.25) {\small\sf Downlink CQI};
\draw[<-,line width=1pt, dashed] (9,1.125) -- (4.5,1.125);
\node[anchor=west] at (4.5,.875) {\small\sf Multiplexing matrix index};
\draw[->,line width=1pt, dashed] (5.,1.125) -- (5., 2.) -- (5.5,2.);
\draw[->,line width=1pt, dashed] (2.,1.25) -- (1.5, 1.25) -- (1.5,5.25) -- (2, 5.25);
\node[anchor=east] at (1.5,5.25) {$\mathbf{R}_{\rm N}^*$};
\draw[->,line width=1pt, dashed] (2.,.75) -- (1.25, .75) -- (1.25,3.25) -- (2, 3.25);
\node[anchor=east] at (1.25,3.25) {$\mathbf{R}_{\rm F}^*$};
\node[anchor=west] at (9.125,.8) {\LARGE\sf \} \small To/from UEs};

\draw[line width=1pt]  (5.5,1.5) rectangle (8.,2.5);
\node[text width=2.cm,text centered] at (6.75,2.) {\small\sf Multiplexing Matrix Library};
\draw[->,line width=1pt, dashed] (6.75,2.5) -- (6.75, 3.);
\node[anchor=west] at (6.75,2.75) {\small\sf Multiplexing Matrix};

\draw[line width=1pt] (10.75,4.75) -- (13.75,4.75) -- (13.75,7.) -- (13.55,7.3) -- (13.75,7.) -- (13.95, 7.3);

\end{tikzpicture}
}
\label{fig:ra-cema-tx-ppt}
}
\subfigure[][Near receiver.]{
\resizebox{.28\hsize}{!}{
	\begin{tikzpicture}[scale=.85]

\draw[rounded corners=10pt, line width=1.5pt] (-.25,0.0) rectangle (6,6.);
\node (tx) at (3.125,5.625) {\bf\textsf{Near receiver}};

\draw[<-,line width=1pt] (.25,3.25) -- (-1.,3.25);
\draw[line width=1pt] (-1.,3.25) -- (-1.,5);
\draw[line width=1pt] (-1.,5) -- (-.75,5.35);
\draw[line width=1pt] (-1.,5) -- (-1.25,5.35);

\node[anchor=north] at (-.75,3.25) {$\mathbf{y}_{\rm N}$};

\draw[line width=1pt]  (.25,1.5) rectangle (1.25,5.);
\node[text width=1.5cm, text centered,rotate=90] (detector) at (.75,3.25)
{\small\textsf{Detector}};
\draw[->,line width=1pt] (1.25,4.5) -- (2.5,4.5);
\node[anchor=south] at (1.875,4.5) {$\lambda_{{\rm F},j}$};
\draw[<-,line width=1pt] (1.25,4.) -- (2.5,4.);
\node[anchor=north] at (1.875,4) {$\mu_{{\rm F},j}$};
\draw[->,line width=1pt] (1.25,2.25) -- (2.5,2.25);
\node[anchor=south] at (1.875,2.25) {$\lambda_{{\rm N},q}$};

\draw[line width=1pt]  (2.5,3.5) rectangle (4.75,5.);
\node[text width=1.9cm, text centered] (enc1) at (3.625,4.25)
{\small\textsf{Far code-word dec. (optional)}};

\draw[line width=1pt]  (2.5,1.5) rectangle (4.75,3.);
\node[text width=2cm, text centered] (enc1) at (3.625,2.25)
{\small\textsf{Near codeword decoder}};
\draw[->,line width=1pt] (4.75,2.25) -- (6.5,2.25);
\node[anchor=south] at (6.5,2.25) {$\mathbf{\hat b}_{\rm N}$};

\draw[line width=1pt, rounded corners=5pt]  (.125,0.25) rectangle (1.875,1.25);
\node[text width=2.5cm, text centered] at (1.,.75) {\footnotesize\textsf{MUX Ma-trix Library}};

\draw[->,line width=1pt, dashed] (.875,1.25) -- (.875,1.5);

\draw[->,line width=1pt, dashed] (3.75,.75) -- (1.875,.75);
\node[anchor=north] at (3.875,.75) {\footnotesize\textsf{MUX mtx index (from TX)}};

\end{tikzpicture}
}
\label{fig:sic-receiver}
}
\caption{Rate-adaptive CEMA system. Solid arrows indicate data signals. Dashed arrows indicate control signals.}
\label{fig:ra-cema-tx-ppt}
\end{figure*}

\section{RA-CEMA Concept}
\label{sec:ra-cema}

A scheme of RA-CEMA transmitter is shown in Fig.~\ref{fig:ra-cema-tx-ppt}.
In general, $U$ UEs experiencing SNR $\rho_u, u=0,\ldots, U-1$,
are served simultaneously using a set of Resource Blocks (RB) each consisting
of a number of time-frequency REs.
Each RB can be allocated for transmission to a single UE or to multiple UEs.
When at least one RB is allocated to more than one UE, the MA scheme
is non-orthogonal.

The RA-CEMA scheduler\footnote{In each transmission interval
a scheduler allocates certain time-frequency-space resources to
a UE which can draw the largest benefits from these particular
resources. However, the scheduler should also ensure that each
UE is served within a certain predetermined delay interval.}
obtains the CQI related to all active users
 and selects for concurrent transmission
users characterized by different CQI values.
We assume that, using this criterion, the scheduler has allocated
a set of RBs, corresponding to
a total number of $G$ REs, to a pair\footnote{In general, RBs can be
allocated to more than two UEs.
Here, for the sake of clarity, we will consider the two-UE case.} of UEs:
a near UE characterized by a good CQI
(high SNR $\rho_{\rm N}$), and
a far UE characterized by a worse CQI (lower SNR
$\rho_{\rm F} < \rho_{\rm N}$).
The channel coefficient for the near (resp. far) UE is
$h_{\rm N}$ (resp. $h_{\rm F}$)
and it is assumed to be constant over a RB.

Similarly as in LTE, using the same criteria that would be used in a
conventional orthogonal MA system the RA-CEMA scheduler computes
a code rate $R_u$ and a modulation order $m_u$ for each UE.
Each information word $\mathbf{b}_u$ is encoded by a channel coding
and rate matching unit, obtaining a codeword $\mathbf{c}_u$ consisting
of $E_u^{(0)} = m_u G$ coded bits.
The total number of coded bits generated by channel coding and rate
matching is therefore
\begin{equation*}
\label{eq:E_TOT}
E_{\rm TOT} = \Sigma_u E_u^{(0)} = G \Sigma_u m_u.
\end{equation*}

In order to accommodate all the $E_{\rm TOT}$
coded bits in the allocated REs,
we apply a \emph{constellation expansion}
approach~\cite{bib:Perot14GCws2},
which consists in
increasing the order of the modulator constellation to a value
\begin{equation*}
m = \Sigma_{u=0}^{U-1} m_u.
\end{equation*}
We obtain an \emph{expanded constellation} $\chi_{\rm EXP}$ 
having size $2^m$ which will be used to transmit the codewords
of all UEs.

Multiplexing of codewords is performed according to a 
\emph{multiplexing matrix} $\mathbf{M}$ of size $m\times G$ 
whose element $(\mathbf{M})_{i,j} \in [0, U-1]$ indicates
the UE whose coded bit is transmitted using the $i$th label
bit of the $j$th constellation symbol.

After multiplexing, a vector $\mathbf{l} = (l_0,...,l_{G-1})$
of $m$-bit
labels is formed and sent to the modulator which performs a
one-to-one
mapping of labels onto complex constellation points thus
forming the transmitted vector $\mathbf{x}$.

\subsection{Design of the Multiplexing Matrices}

On the degraded BC the near user is supposed to be
able (due to its higher SNR) to perfectly decode any codeword
transmitted to the far user, allowing in that way its receiver 
to perfectly remove the far user interfering signal.
If superposition coding (SC)~\cite{bib:CoverThomas} is used,
such perfect removal is made possible by allocating a larger
power to the far user signal than to the near user signal,
assuming that both far and  near user codewords use the same
number of modulation symbols.

In RA-CEMA, where the near- and far-UE codewords  are 
multiplexed onto common modulation symbols, in general it is not
feasible to set arbitrary powers to the coded bits of the multiplexed users.
Thus, in  order to make the transmitted signal energy of the far user
larger than for the near user, we use a combination of two
techniques: A) special usage of unequal bit-level capacities
in the modulation constellation binary labels; and B)
unequal code word lengths.

The concept of bit-level capacity has been introduced
in~\cite{bib:Stier10IZSC} and corresponds to the mutual information
of each bit in a constellation binary label. Bits occupying different 
positions in the label exhibit different capacities which depend
on the shape of the constellation
and on the specific binary labeling.
In the RA-CEMA multiplexer, each row of
 $\mathbf{M}$ corresponds to a different
position in the constellation label.
Therefore, all bits in the same row exhibit the same 
bit-level capacity, whereas bits in different rows  possibly
exhibit different bit-level capacities.
We arbitrarily choose to associate label bits with a higher
capacity to the first rows of $\mathbf{M}$ 
and label bits with lower capacities to  other rows in
non-increasing order of capacity.
As a higher energy per codeword results in a higher transmission
rate, assigning the label bits with higher capacities to the far-UE 
codeword achieves the same effect as  
allocating a larger power to the far-UE signal. 

In orthogonal MA, all REs in a RB are allocated to only one UE.
In RA-CEMA, it is still possible to have some REs entirely
allocated to a single UE.
However, when most of the REs in a RB are allocated to a
single UE, the multiplexing scheme becomes similar to
an orthogonal scheme and therefore little rate gains
with respect to time sharing are expected.
We conclude that, by minimizing the number of REs allocated
to a single UE, we obtain MA schemes with higher gains.

In summary, the multiplexing matrix is designed according to
the following principles:
\begin{enumerate}
\item Assign label bits with higher capacity to the far-UE codeword.
\item Maximize the number of REs having their label bits
assigned to multiple UE codewords.
\end{enumerate}

An example of multiplexing matrix designed according to these
principles is 
\begin{equation*}
\resizebox{.28\textwidth}{!}{$
\mathbf{M} =
\left[
\begin{tabular}{ccccccc}
{\rm F}& ... & {\rm F}& {\rm F} & ... & {\rm F} \\
{\rm F}& ... & {\rm F}& {\rm F} & ... & {\rm F} \\
{\rm F}& ... & {\rm F}& {\rm N} & ... & {\rm N} \\
{\rm N}& ... & {\rm N}& {\rm N} & ... & {\rm N} \\
{\rm N}& ... & {\rm N}& {\rm N} & ... & {\rm N} \\
{\rm N}& ... & {\rm N}& {\rm N} & ... & {\rm N}
\end{tabular}\right]
$}
\end{equation*}
where, for the sake of clarity, user indices $u \in \{0, 1\}$ have been
replaced by  tags $\{\rm N,F\}$. Here the order of the expanded 
constellation is $m=6$, hence a constellation with size $2^6$
like 64-QAM could be used.

For some constellations, multiple label bits are characterized
by the same bit-level capacity. In $M$-QAM, for example, each capacity
level is common to two label bits. In such cases, different multiplexing
matrices might be equivalent in terms of performance. 
To clarify this, consider the following matrix:
\begin{equation*}
\resizebox{.28\textwidth}{!}{$
\mathbf{M}^\dagger = \left[
\begin{tabular}{ccccccc}
{\rm F}& ... & {\rm F}& {\rm F} & ... & {\rm F} \\
{\rm F}& ... & {\rm F}& {\rm F} & ... & {\rm F} \\
{\rm N}& ... & {\rm N}& {\rm N} & ... & {\rm N} \\
{\rm F}& ... & {\rm F}& {\rm N} & ... & {\rm N} \\
{\rm N}& ... & {\rm N}& {\rm N} & ... & {\rm N} \\
{\rm N}& ... & {\rm N}& {\rm N} & ... & {\rm N}
\end{tabular}\right].
$}
\end{equation*}
When used with 64-QAM, matrices $\mathbf{M}^\dagger$ and
$\mathbf{M}$ are equivalent because their third row and fourth row
correspond to label bits characterized by  the same capacity level.

In order to further enhance flexibility in controlling the transmitted 
signal energies of the multiplexed users, we let the \emph{actual 
codeword length} $E_{\rm F}$ of the \emph{far} user to be proportional to the 
targeted Spectral Efficiency (SE) of that user. For example, if the targeted SE is close to its 
single-user SE, then the codeword length should be almost equal to 
$mG$. On the other side, if the targeted SE of the \emph{near} user is close to 
its single user SE, then the codeword length of the far user is close to zero. 
Different far user codeword lengths produce different multiplexing 
matrices.

Using the described design procedure, a concrete example of
matrix library has been designed for a system with SNR values
$\rho_{\rm N} = 12$ dB and $\rho_{\rm F} = 6$ dB.
In this case, we have $m_{\rm F} = 2$ and $m_{\rm N} = 4$,
therefore the expanded constellation has order $m=6$ (64-QAM).
The number of available REs is $G=240$ and the total number of
coded bits is $E_{\rm F} + E_{\rm N} = mG = 1440$. 
The matrix library is given in Tab.~\ref{tab:mtxlib} as the set of
matrices $\{\mathbf{M}_h\}, h = 0,\ldots, 8$.
Such library will be used later in Sec.~\ref{sec:perf-eval} for 
performance evaluation.

\begin{table}[!ht]
\centering
\caption{Parameter values of the multiplexing matrix library designed for $\rho_{\rm N} = 12\;\rm dB$, $\rho_{\rm F} = 6\;\rm dB$ and $G=240$ REs.}
\begin{tabular}{cc|cc|cc}
Matrix ID & $E_{\rm F}$ & Matrix ID & $E_{\rm F}$ & Matrix ID & $E_{\rm F}$\\
\hline \hline
$\mathbf{M}_0$ & 480 & $\mathbf{M}_3$ & 640   & $\mathbf{M}_6$ & 80\\
$\mathbf{M}_1$ & 400 & $\mathbf{M}_4$ & 240  & $\mathbf{M}_7$ &  840\\
$\mathbf{M}_2$ & 320 & $\mathbf{M}_5$ & 160 & $\mathbf{M}_8$ & 960
\end{tabular}
\label{tab:mtxlib}
\end{table}

\subsection{Optimization of Information Word Lengths}
\label{sec:optproc}

The capacity region of RA-CEMA is evaluated by using two alternative measures:
the Modulation and Coding Scheme (MCS) rate and the spectral efficiency
of each user.
We define the MCS rate $R_u^{(\rm MCS)}$ as 
\begin{equation}
\label{eq:R_MCS}
R_u^{(\rm MCS)} = R_u m_u = K_u / G \quad \rm [inform.\;bits/symbol]
\end{equation}
where  $R_u$ is the code rate, $m_u$ is the modulation order and $K_u$ is
the information word length.
The MCS rate is a kind of generalization
of single-user code rate reflecting the impact of the modulation order
to the number of information bits transmitted per modulation symbol.
The reason for, in network information theory, only the code
rate $R_u$ is used as the basic measure for defining the rate regions
of multiuser
channels~\cite{bib:ElGamalKim} is that the modulation is typically ignored.

Using the BLock Error Rate (BLER) obtained by simulation, we
estimate the spectral efficiency ${\rm SE}_u$ as
\begin{equation}
\label{eq:SE}
{\rm SE}_u(K_u; \mathbf{M}) = \left[1-{\rm BLER}(K_u)\right] R_u^{(\rm MCS)}\; \rm [bits/s/Hz].
\end{equation}
This definition of SE combines the MCS rate with BLER, reflecting in that way 
the degree to which a certain MCS rate is achievable. Therefore the SE may
be considered a more realistic measure for determining the capacity region.

The achievable MCS rate pairs and SE pairs are
obtained through two different optimization procedures.
The first procedure (proc. 1) maximizes the aggregate SE
defined as
\begin{equation}
\label{eq:SEagg}
{\rm SE}_{\Sigma}(K_{\rm N}, K_{\rm F}; \mathbf{M}) =
\Sigma_{u\in\{{\rm N},{\rm F}\}}{{\rm SE}_u(K_u; \mathbf{M})}.
\end{equation}
Using (\ref{eq:SEagg}), we obtain the optimum pair of information
word lengths
\begin{equation}
\label{eq:opt1}
(K_{\rm N}^*, K_{\rm F}^*) = \arg\max_{K_{\rm N},
K_{\rm F}}{\rm SE}_{\Sigma}(K_{\rm N}, K_{\rm F}; \mathbf{M}).
\end{equation} 
The corresponding pair of rates $(R_{\rm N}^{(\rm MCS)}, R_{\rm F}^{(\rm MCS)})$ or spectral efficiencies $({\rm SE}_{\rm N}, {\rm SE}_{\rm F})$ are obtained from 
(\ref{eq:R_MCS}) and (\ref{eq:SE}). The corresponding code rates
are computed as $R_{\rm N}^* = K_{\rm N}^* / E_{\rm N}$ and 
$R_{\rm F}^* = K_{\rm F}^* / E_{\rm F}$

The second procedure (proc. 2) aims at maximizing 
the aggregate MCS rate
\begin{equation*}
R_{\Sigma}^{(\rm MCS)}(K_{\rm N}, K_{\rm F}; \mathbf{M}) =
\Sigma_{u\in\{{\rm N},{\rm F}\}}{R_u^{(\rm MCS)}(K_u; \mathbf{M})}
\end{equation*}
subject to the link quality constraint ${\rm BLER}(K_u) < \epsilon, \forall u.$
According to this criterion, the optimal pair of information word lengths
is obtained as
\begin{equation}
\label{eq:opt2}
(K_{\rm N}^+, K_{\rm F}^+) = \arg\max_{K_{\rm N}, K_{\rm F}}R_{\Sigma}^{(\rm MCS)}(K_{\rm N}, K_{\rm F}; \mathbf{M})
\end{equation}
and the corresponding pair of SE $({\rm SE}_{\rm N}, {\rm SE}_{\rm F})$ and
pair of MCS rates $(R_{\rm N}^{(\rm MCS)}, R_{\rm F}^{(\rm MCS)})$ are obtained from 
(\ref{eq:R_MCS}) and (\ref{eq:SE}).

\subsection{Interference Cancellation (IC) receiver for RA-CEMA}

The near receiver performs IC as shown in Fig.~\ref{fig:sic-receiver}.
The detector computes the log-likelihood ratios (LLRs) of symbols $s_n$ of the
expanded constellation $\chi_{\rm EXP}$ as
\begin{equation}
\label{eq:sym_llrs}
\lambda_{n,k}=\log\frac{P(x_k=s_n | y_{{\rm N},k})}{P(x_k=s_0|y_{{\rm N},k})},
\; n=0,\ldots,|\chi_{\rm EXP}|-1 
\end{equation}
where $k\in\{0, \ldots, G-1\}$ is the time index, $x_k$ is the symbol transmitted
at time $k$ and $y_{{\rm N},k}$ is the complex
sample received by  the near user at time $k$.
The detector then computes the binary LLRs of
codeword $\mathbf{c}_{\rm F}=(c_{{\rm F},0},\ldots,c_{{\rm F},E_{\rm F}-1})$ as
\begin{eqnarray}
\label{eq:cbit_llrs}
\lambda_{{\rm F},j}  &= &\log\frac{P(c_{{\rm F},j}=1|\mathbf{y}_{{\rm N}})}{P(c_{{\rm F},j}=0|\mathbf{y}_{{\rm N}})} \\
& = & \mathop{\mathrm{max}^*}_{n:\mathcal{L}_{\xi_{\rm F}(j)}(s_n)=1}\lambda_{n,\omega_{\rm F}(j)} \nonumber
- \mathop{\mathrm{max}^*}_{n:\mathcal{L}_{\xi_{\rm F}(j)}(s_n)=0}\lambda_{n,\omega_{\rm F}(j)}
\end{eqnarray}
where $\omega_{\rm F}(j)\in\{0, \ldots, G-1\}$ indicates the symbol
in which bit $j$ of $\mathbf{c}_{\rm F}$ has been  transmitted
and $\xi_{\rm F}(j)\in\{0, \ldots, m-1\}$ indicates its position in the
binary label.
Here, $\mathcal{L}_\xi(s_n)$ indicates the value of bit $\xi$ in the binary
label associated to constellation symbol $s_n$ and
$\max{}^*(a, b) \triangleq \log(e^a+e^b)$.

The computed LLRs are sent to the far codeword decoder which computes
updated \emph{a-posteriori} extrinsic LLRs $\mu_{{\rm F},j}$ of coded bits.
Such updated LLRs are fed back to the detector and used as \emph{a-priori} information
of the far-codeword bits.

The detector updates
the LLRs of constellation symbols as
\begin{equation*}
\bar\lambda_{n,k} = \lambda_{n,k} + \Sigma_{i=0}^{m-1}{\mathcal{L}_i(s_n) \mu_{v_{ik},z_{ik}}}
\end{equation*}
with $v_{ik} \in \{\rm N, F\}$,
$\mu_{{\rm N}, z} \equiv 0, \forall z$ and $z_{ik} \in [0, \ldots, E_{v_{ik}}-1]$
and computes binary LLRs of codeword
\mbox{$\mathbf{c}_{{\rm N}}$} as
\begin{eqnarray*}
\lambda_{{\rm N},q}  &= &\log\frac{P(c_{{\rm N},q}=1|\mathbf{y}_{{\rm N}})}{P(c_{{\rm N},q}=0|\mathbf{y}_{{\rm N}})} \\
& = & \mathop{\mathrm{max}^*}_{n:\mathcal{L}_{\xi_{\rm N}(q)}(s_n)=1}\bar\lambda_{n,\omega_{\rm N}(q)} 
- \mathop{\mathrm{max}^*}_{n:\mathcal{L}_{\xi_{\rm N}(q)}(s_n)=0}\bar\lambda_{n,\omega_{\rm N}(q)}.
\end{eqnarray*}
These LLRs are sent to the near codeword decoder which 
computes the estimate
$\hat{\mathbf{b}}_{{\rm N}}$.

At the far receiver, the detector computes LLRs on the transmitted symbols
$\lambda_{n,k}$ as in (\ref{eq:sym_llrs}) with $y_{{\rm N},k}$ replaced by $y_{{\rm F},k}$.
LLRs of the far codeword bits are
computed as in (\ref{eq:cbit_llrs}) with $\mathbf{y}_{{\rm N}}$ replaced by
$\mathbf{y}_{{\rm F}}$.
Finally, the far codeword decoder applies the code constraints and
computes the estimate $\hat{\mathbf{b}}_{{\rm F}}$.

The near- and the far-codeword decoders are iterative turbo decoders.
They compute \emph{a-posteriori} extrinsic LLRs of coded bits
and of information bits
by iterative execution of the soft-in soft-out (SISO)
algorithm~\cite{bib:BenDivMonPol98a}.

\begin{table}[!b]
\centering
\caption{Selected LTE modulation and coding schemes.}
\resizebox{.9\hsize}{!}{
\begin{tabular}{c@{ }c@{ }r@{ }c@{ }r@{.}l||c@{ }c@{ }r@{ }c@{ }r@{.}l}
Index &	$m_u$	& $K_u$ & $R_u^{(\rm MCS)}$ & \multicolumn{2}{p{1.2cm}}{$\rho_{\min}$ [dB]}	&
Index	& $m_u$	& $K_u$ & $R_u^{(\rm MCS)}$ & \multicolumn{2}{p{1.2cm}}{$\rho_{\min}$ [dB]} \\     
\hline
1	& 2  & 96 	&	0.4	 	& -2&8	& 18	& 4 	&	368	& 1.533	& 4&7 \\
2	& 2	& 112	& 0.467	& -2&1	& 19	& 4	& 384	& 1.6		& 5&0 \\
3	& 2	& 128	& 0.533	& -1&6	& 20	& 4	& 400	& 1.667	& 5&3 \\
4	& 2	& 144	& 0.6		& -1&1	& 21	& 4	& 416	& 1.733	& 5&5 \\
5	& 2	& 160	& 0.667 	& -0&8	& 22	& 4	& 432	& 1.8		& 5&8 \\
6	& 2  & 176	& 0.733	& -0&3	& 23	& 4	& 448	& 1.867	& 6&1 \\
7	& 2	& 192	& 0.8		& 0&2	& 24	& 4	& 464	& 1.933	& 6&4 \\
8	& 2	& 208	& 0.867	& 0&7	& 25	& 4	& 480	& 2			& 6&6 \\
9	& 2	& 224	& 0.933	& 1&1	& 26	& 4	& 512	& 2.133	& 7&1 \\
10 & 2	& 240	& 1			& 1&5	& 27	& 4	& 544	& 2.267	& 7&7 \\
11& 2	& 	256	& 1.067	& 	2&0	& 28	& 4	& 576	& 2.4		& 8&2 \\
12& 2	& 	272	& 1.133	& 2&3	& 29	& 4	& 608	& 2.533	& 8&8 \\
13& 2	& 	288	& 1.2		& 2&8	& 30	& 4	& 640	& 2.667	& 9&3 \\
14& 2	& 	304	& 1.267	& 3&2	& 31	& 4	& 672	& 2.8		& 9&8 \\
15& 2	& 	320	& 1.333	& 3&6	& 32	& 4	& 704	& 2.933	& 10&3 \\
16& 2	& 	336	& 1.4		& 4&0	& 33	& 4	& 736	& 3.067	& 10&8 \\
17&  2 & 	352	& 1.467	& 	4&4	& 34	& 4	& 768	& 3.2		& 11&3
\end{tabular}}
\label{tab:mcs}
\end{table}

\begin{figure}[!ht]
\centering
\subfigure[][Spectral efficiency pairs.]{
	\includegraphics[scale=.36,clip=true,trim=4.2cm 2.7cm 4.3cm 2.7cm]{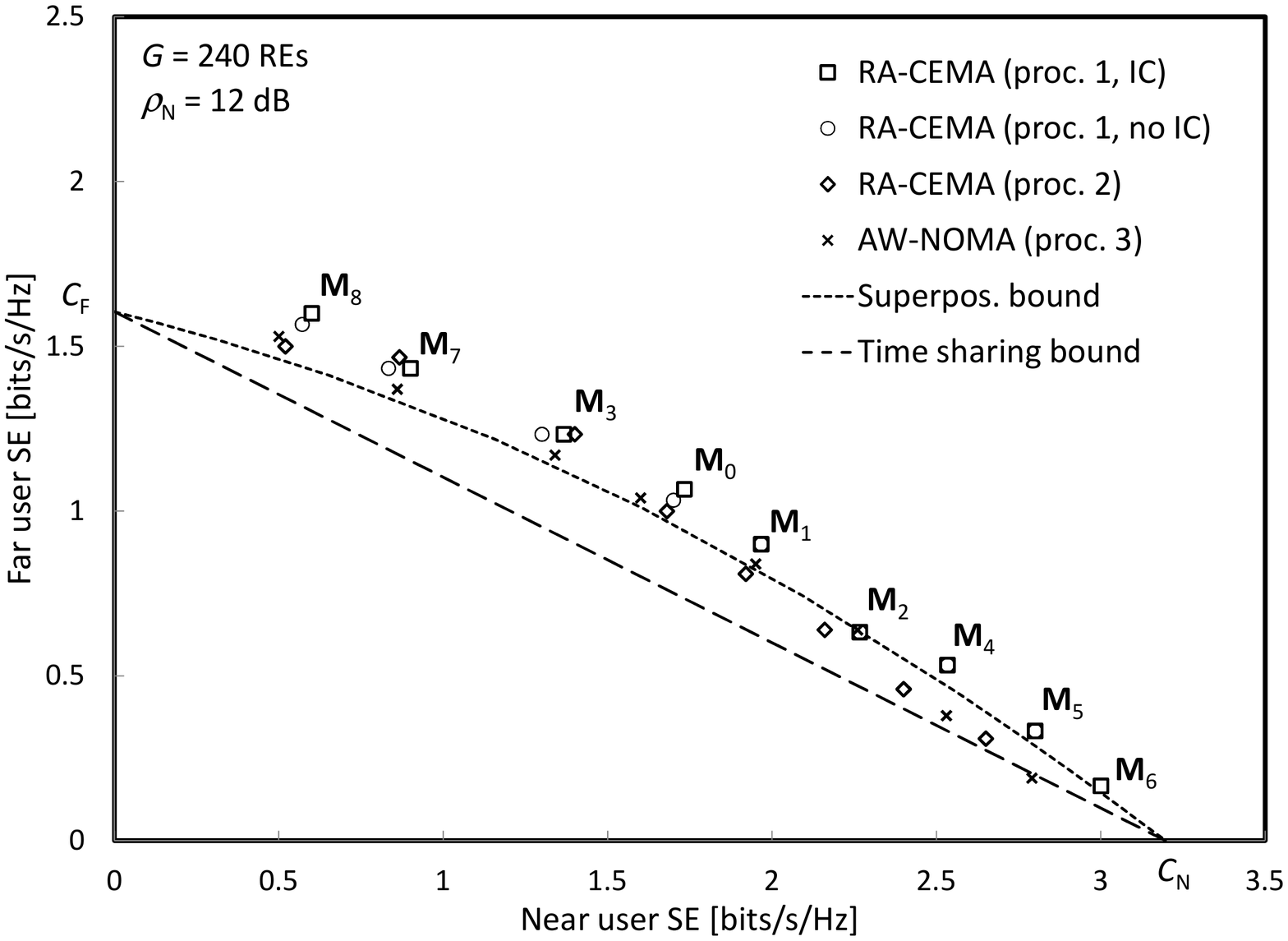}
	\label{fig:ra-cema-se-awgn}
}
\subfigure[][MCS rate pairs.]{
\includegraphics[scale=.36,clip=true,trim=3.9cm 2.7cm 4.cm 2.7cm]{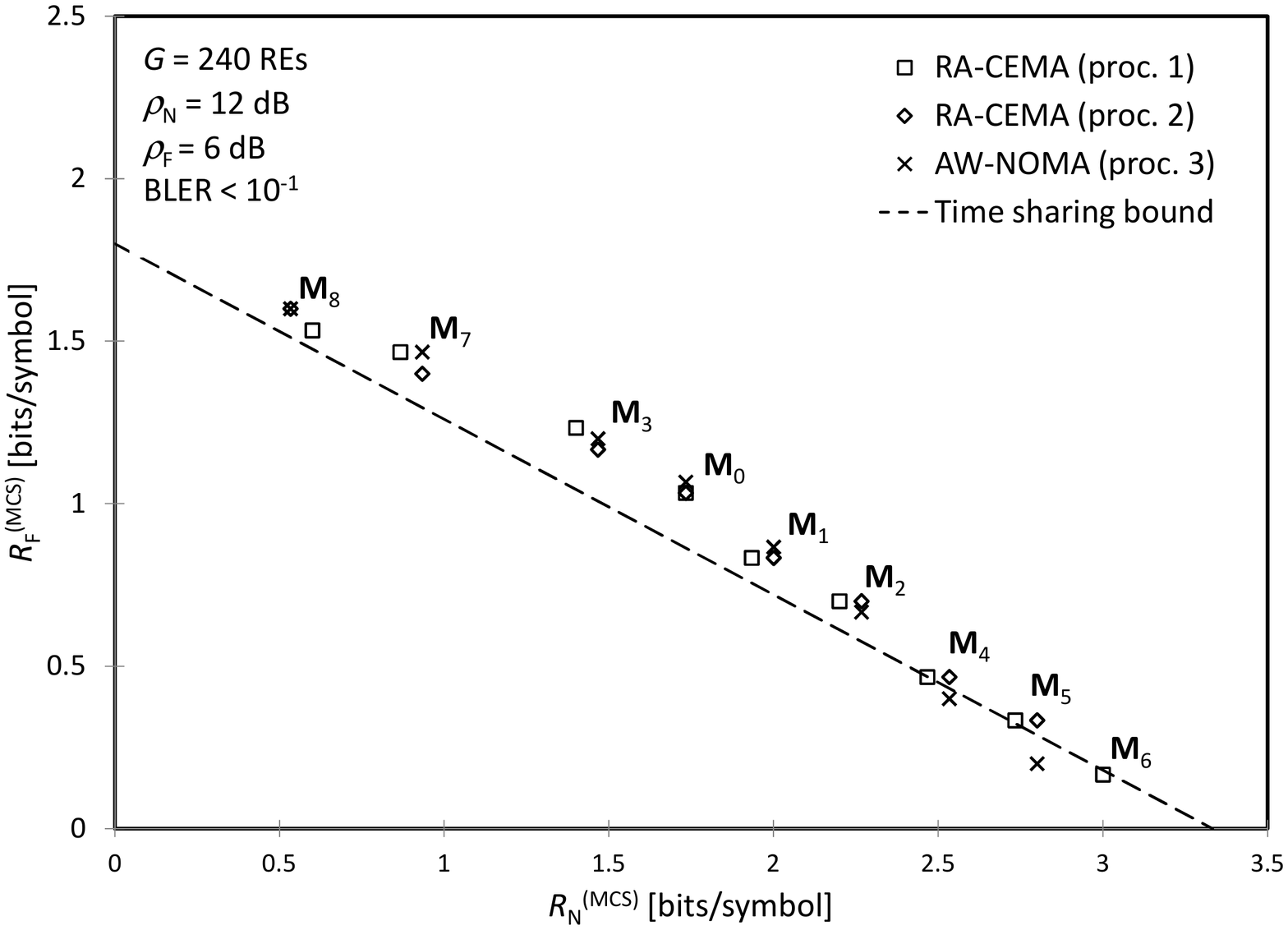}
\label{fig:ra-cema-rates-awgn}
}
\caption{Capacity region of RA-CEMA and AW-NOMA on the degraded BC
with AWGN.}
\label{fig:ra-cema-awgn}
\end{figure}

\section{Performance Evaluation}
\label{sec:perf-eval}

Fig.~\ref{fig:ra-cema-awgn} shows the capacity region of RA-CEMA
and of AW-NOMA on the degraded BC with AWGN
obtained for two users experiencing SNRs $\rho_{\rm N} = 12$ dB
and $\rho_{\rm F} = 6$ dB.
The standard LTE turbo code, rate-matching scheme and QAM
constellations with Gray labelling~\cite{bib:LTEr12-212} have
been used.
In simulations, the turbo decoder performs $N_{\rm IT} = 10$
iterations.

The time sharing bound corresponds to
the achievable pairs of SEs or MCS rates with orthogonal multiplexing.
The single-user SEs for the near UE is $C_{\rm N} = 3.2$ bits/s/Hz,
while for the far user we have $C_{\rm F} = 1.6$ bits/s/Hz).
By allocating non-overlapping sub-intervals of different duration
to the two users, it is possible to achieve all the rate pairs 
on the line connecting the points $(C_{\rm N}, 0)$ and $(0, C_{\rm F})$.

In Fig.~\ref{fig:ra-cema-awgn} we also plot an approximate result labelled 
``Superposition bound'' used to predict the achievable rate pairs based on
the single-user capacities  $C_{\rm N}$ and
$C_{\rm F}$ achieved by the two users when transmitting alone on the
AWGN channel.
Applying the inverse of the AWGN capacity function  $\mathscr{C}^{-1}(y)$
(where $y = \mathscr{C}(x) \triangleq \log_2(1+x)$), we obtain 
\begin{equation*}
\tilde{\rho}_{\rm N} = \mathscr{C}^{-1}(C_{\rm N}) \simeq 9.15 {\rm dB}; \quad
\tilde{\rho}_{\rm F} = \mathscr{C}^{-1}(C_{\rm F}) \simeq 3.1 \rm dB.
\end{equation*}
Finally, we apply the boundary equations of the SC rate region
\begin{equation*}
R_{\rm N} = \mathscr{C}(\alpha \tilde{\rho}_{\rm N}); \quad R_{\rm F} =
\mathscr{C}\left(\frac{(1-\alpha) \tilde{\rho}_{\rm F}}{\alpha \tilde{\rho}_{\rm F}+1}\right)
\end{equation*}
where $\alpha\in[0,1]$ and we obtain the curve labelled ``Superpos. bound'' in
Fig.~\ref{fig:ra-cema-awgn}.
This bound fairly accurately predicts the actual 
boundary of the capacity region, therefore it can be considered as a useful
design tool. Moreover, we observe that RA-CEMA exhibits relevant SE and
rate improvements with respect to time sharing.

In order to compare the performance of RA-CEMA with the
previously proposed
AW-NOMA scheme, the optimization procedure 
described in~\cite[Sec.~V-B]{bib:Vanka12TranWC} (denoted as
``proc. 3'' in Fig.~\ref{fig:ra-cema-awgn}) 
has been applied to determine
the rate pairs corresponding to certain values of power ratios $\alpha$
and $(1- \alpha)$ allocated to near and far UE respectively. 
However, since the MCSs therein
considered are different from those used here, the step size for
$\alpha$ has been adjusted\footnote{We
observed a correlation between the step size and the ``granularity'' of the
MCS set. In particular, the step size for $\alpha$ should be less than the
(approximate) SNR step size computed on column $\rho_{\min}$ of
Tab.~\ref{tab:mcs}.}  to 0.1 dB.
Tab.~\ref{tab:mcs} shows the used MCS parameters and
the SNR $\rho_{\min}$ needed to achieve BLER
below $\epsilon = 10^{-1}$ with single-user
transmission on the AWGN channel.
Another set of SNR values,
herein omitted for lack of space, has been
obtained for the fading channel. 

In Fig.~\ref{fig:ra-cema-se-awgn}, results are expressed in terms of
SE pairs $({\rm SE}_{\rm N}, {\rm SE}_{\rm F})$
computed using (\ref{eq:opt1}) and (\ref{eq:opt2}).
Using the same two equations, the MCS rate pairs
$(R_{\rm N}^{(\rm MCS)}, R_{\rm F}^{(\rm MCS)})$ shown in
Fig.~\ref{fig:ra-cema-rates-awgn} have been found.
AW-NOMA SE and rate pairs are computed using the procedure
mentioned above.

Fig.~\ref{fig:ra-cema-se-awgn} also shows the SE pairs achieved 
when the near-user receiver does not perform IC (points labelled
``RA-CEMA (proc. 1, no IC)'').
RA-CEMA without IC in the near-user receiver shares some basic
features of the Bit-Interleaved Coded Modulation (BICM) transmission
with Gray labelling ~\cite{bib:BICM}.
It has been observed in~\cite{bib:SongTVT14} that the label bits of
BICM with Gray labelling are almost independent. From this
observation it can be concluded that even the exact knowledge of
the far-user coded bits should be of little help to the near-user receiver.
Hence, removing IC from the near-user receiver should not make a
significant deterioration of the near-user achieved SE.
This conclusion has been confirmed by the simulation results
in Fig.~\ref{fig:ra-cema-se-awgn}.

\begin{figure}[!t]
\centering
\includegraphics[scale=.37,clip=true,trim=3.5cm 2.7cm 4.cm 2.7cm]{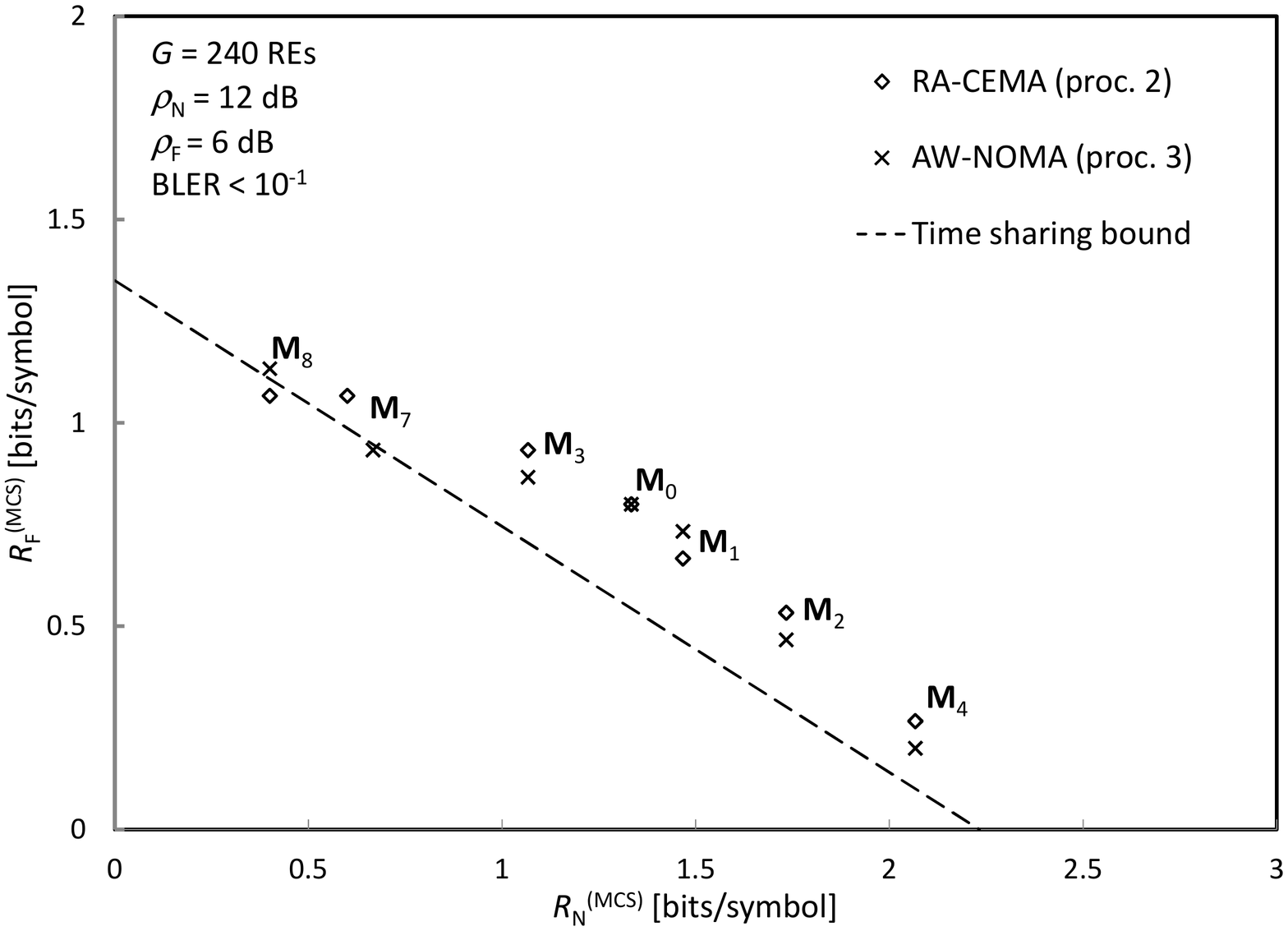}
\caption{Rate region of RA-CEMA and AW-NOMA on the degraded BC
with block fading expressed through MCS rate pairs.}
\label{fig:ra-cema-rates-fading}
\end{figure}

Similar results are shown in Fig.~\ref{fig:ra-cema-rates-fading}. 
Here, block fading with coherence time of 80 symbols (approx.
equal to the size of one LTE RB) has been considered as an
additional channel impairment.
A channel interleaver of size $G$ connected to the modulator output
has been employed to de-correlate fading within each block.
Also on the fading channel, both RA-CEMA and AW-NOMA
perform better than time sharing.

\section{Conclusions}
\label{sec:conclusions}

A new multiple access scheme for the degraded broadcast
channel, which performs
multiplexing of UE signals by storing their codewords
into a multiplexing matrix and then maps the columns
onto constellation symbols, has been proposed.
The new scheme has similar performance
as conventional schemes based on superposition coding
while avoiding their potential standardization problems
caused by the use of unconventional constellations.


\bibliographystyle{IEEEtran}
\bibliography{IEEEabrv,../xbib}

\end{document}